\documentclass[10pt]{article}
\usepackage{amsmath}
\usepackage[dvips]{graphicx}

\begin{document}

\begin{center}
\textbf{\large Minimum Length Cutoff in Inflation and Uniqueness of the
Action}\\[0pt]
\vspace{48pt}

\textbf{A. Ashoorioon\footnote{%
amjad@astro.uwaterloo.ca}, A. Kempf\footnote{
akempf@uwaterloo.ca}, R.B. Mann\footnote{%
mann@avatar.uwaterloo.ca} }

\textbf{\textrm{\vspace{12pt} \centerline{Departments of Physics and
Applied Mathematics, University of Waterloo and} \centerline{Perimeter
Institute for Theoretical Physics}
 \centerline{ Waterloo,
Ontario, N2L 3G1, Canada} }}
\end{center}

\begin{equation*}
\end{equation*}

%\hfill{} \vspace{32pt}
According to most inflationary models, fluctuations that are of
cosmological size today started out much smaller than any plausible cutoff
length such as the string or Planck lengths. It has been shown that this
could open an experimental window for testing models of the short-scale
structure of space-time. The observability of effects hinges crucially,
however, on the initial conditions imposed on the new comoving modes which
are continually being created at the cutoff length scale. Here, we address
this question while modelling spacetime as obeying the string and quantum
gravity inspired minimum length uncertainty principle. We find that the
usual strategy for determining the initial conditions faces an unexpected
difficulty because it involves reformulating the action and discarding a
boundary term: we find that actions that normally differ merely by a
boundary term can differ significantly when the minimum length is
introduced. This is possible because the introduction of a minimum length
comes with an ordering ambiguity much like the ordering ambiguity that
arises with the introduction of $\hbar $ in the process of quantization.

\addtolength{\baselineskip}{1.4mm} \thispagestyle{empty}

%\vspace{48pt}

\setcounter{footnote}{0}

\section{Introduction}

Some of the predictions of fundamental theories of physics can only be
observed on energy scales as high as the Planck scale. The availability of
such high energies in the early universe and the huge separation between
conventional accelerator experiments and the Planck scale has led many to
turn from accelerator-based experiments to cosmological observations in
order to test such theories. Inflationary cosmology \cite{1} is one of the
paradigms that may serve this purpose. There, it is assumed that quantum
fluctuations of the inflaton are stretched by inflationary expansion to
cosmological scales. About 60 e-folds of inflationary expansion are
necessary to solve many of the puzzles of big bang cosmology but in most
inflationary models the expansion is much larger. In models proposed by
Linde \cite{2} the universe has expanded by a factor of $10^{10^{12}}$. An
implication of these models is, therefore, that our observable universe was
of \textit{sub-Planckian} size at the beginning of the (last) inflationary
period. This suggests that inflation could act as a magnifying glass for
probing the short distance structure of space-time.

A similar question had arisen concerning a possible sensitivity of black
hole radiation to transplanckian physics. There, it has been found that
Hawking radiation is largely immune to transplanckian effects, see e.g. \cite%
{brout-bh}. In the case of inflation, however, it has been found that the
inflationary predictions for the cosmic microwave background (CMB) do
possess a small and possibly even observable sensitivity to modifications of
quantum field theory in the ultraviolet. To this end, various examples of
ultraviolet-modified dispersion relations, some motivated by solid state
analogs, have been tested for their effects on inflation, see \cite{3,4,5,6}%
. In particular, and this will be our interest here,  the ultraviolet cutoff
described by a lower bound in the formal uncertainty in position, $\Delta
x_{min}$, has also been investigated for its implications in inflation, see %
\cite{7,20,10,11}.

To model the small scale structure of space through a finite minimum
position uncertainty $\Delta x_{min}$ is of interest because the
corresponding modified uncertainty principle has been motivated to arise
from quite general quantum gravity arguments as well as from string theory,
see e.g. \cite{8,9,akucr,KMM}. In fact, any theory with this type of
ultraviolet cutoff can be written, equivalently, as a continuum theory and
as a lattice theory, see \cite{akl}. While in the continuum formulation the
theory displays unbroken external symmetries, the theory's ultraviolet
regularity is displayed in its lattice formulation.

Indeed, it has been found that inflationary predictions for the CMB are
sensitive to the natural ultraviolet cutoff if the cutoff is modelled
through a finite minimum uncertainty in positions, $\Delta x_{min}$. The
magnitude by which the cutoff affects the predicted scalar and tensor
spectra in the CMB was found to depend crucially on the initial conditions
when a mode's evolution begins, which is when its proper wave length is the
minimum length. These initial conditions determine how close the modes'
state is to the adiabatic vacuum during the period of adiabatic evolution
before the mode crosses the Hubble horizon. If the modes are in the
adiabatic vacuum during the phase of adiabatic evolution then the effects of
Planck scale physics on inflationary predictions should be no bigger than of
the order of $\sigma ^{2}$, see \cite{20}, where:
\begin{equation}
\sigma =\frac{\Delta x_{min}}{L_{Hubble}}
\end{equation}%
Here, $L_{Hubble}$ is the Hubble length during inflation. Thus, we have
approximately $\sigma \approx 10^{-3}$ if the cutoff length, $\Delta x_{min}$%
, is at the string scale and $\sigma \approx 10^{-5}$ if the cutoff length
is at the Planck scale. In principle, however, the cutoff can lead to
arbitrarily large effects, namely if the modes' state during the adiabatic
phase differs strongly from the adiabatic vacuum (see also \cite{pad}). In
this case, the modulus of the mode functions oscillates at horizon crossing
and these oscillations translate into characteristic oscillations in the CMB
spectra. This possibility is restricted, however, by the need to keep the
back-reaction small \cite{tanaka}. Interestingly, Easther \textit{et.al. }%
\cite{10,11} found that this constraint still allows nontrivial vacua with
effects as large as of order $\sigma $. Effects of this magnitude might
reach the threshold of observability.

So far, initial conditions have been proposed based on analyticity arguments %
\cite{20} and based on similarity to the Bunch Davies vacuum \cite{10,11}. A
further suggestion is to minimize the field uncertainties \cite%
{danielsson,greene3}. Still, however, the crucial question how to determine
initial conditions for the new comoving modes that are continually being
created during an expansion has not been conclusively answered. The problem
is of course equivalent to identifying the vacuum state.

Here, we address this problem by reconsidering how the vacuum state is
usually identified within inflationary QFT without a minimum length.
Namely, the usual strategy is to make use of the fact that the action can
be rewritten so as to resemble the familiar action of a field on Minkowski
space with time-dependent mass term. When quantizing, one then chooses the
vacuum as one does for Minkowski space theories. We will find that this
method is no longer reliable when there is a minimum length. The reason is
that the reformulation of the action requires the neglect of a boundary
term that ceases to be a boundary term once the minimum length is
introduced. We find that the differences are small but noticeable both in
the initial conditions and in the evolution equations. {This shows that in
any approach to introducing a minimum length into inflation this will have
to be taken into account: reformulations of an action that appear to be
harmless due to neglect of a boundary term can lead to an unintended
modification of the theory.}

To see how this phenomenon can arise, let us recall that the particular
model of a natural ultraviolet cutoff that we are considering is described
by quantum mechanical uncertainty relations with correction terms in the
ultraviolet, of the form
\begin{equation}
\Delta x \Delta p \ge \frac{\hbar}{2}\left(1+\beta~(\Delta p)^2 + ...\right)
\label{1a}
\end{equation}
where $\beta >0$ is a positive constant. In the simplest case, such an
uncertainty relation arises from the modified commutation relation:
\begin{equation}
\lbrack \mathbf{X},\mathbf{P}]=i\hbar (\mathbf{1}+\beta ~\mathbf{P}^{2})
\label{1}
\end{equation}
It is not difficult to show that the uncertainty relation then implies a
finite lower bound to the position uncertainty $\Delta x$:
\begin{equation}
\Delta x_{min} = \hbar \sqrt{\beta}
\end{equation}
By choosing $\beta$ appropriately we obtain a cutoff at the string or at the
Planck scale. This type of ultraviolet cutoff was introduced into quantum
field theory in \cite{akf} and then into inflationary cosmology in \cite{7}.

It is clear that similar to quantization, which changes the commutativity
properties and therefore comes with a well-known ordering ambiguity, the
introduction of a minimum length through an equation such as Eq. (\ref{1})
changes the commutativity properties and therefore comes with an ordering
ambiguity. In principle, of course, ordering ambiguities can be of arbitrary
magnitude. For example, a classical system is unchanged by adding terms of
the form $(xp-px)f(x,p)$ to its Hamiltonian $H$. When promoting  $x$ and $p$
to operators the resulting terms become proportional to $\hbar $ and could
be arbitrarily large and significant to the evolution. In the case we
consider here, normally vanishing terms of the form $(xp-px-i\hbar )g(x,p)$
similarly become nonzero when $\beta \neq 0$. Here, the new Hamiltonian is
determined only up to terms that vanish when setting the minimum length to
zero. Those terms can be arbitrarily large and, in principle, only
experiments could decide which choice is correct. This is to be expected in
any approach to introducing some form of a natural minimum length.

Of course, in the case of quantization it has proven to be a very reliable
strategy to adopt the minimalist approach to resolving the ordering
ambiguity: write the Hamiltonian in its most simple and symmetric form and
leave it unchanged when introducing $\hbar$, i.e. do not introduce terms by
hand. The same minimalist approach has tacitly been applied in the
literature when the minimum length uncertainty relation has been used in
inflationary quantum field theory. We will now review this procedure,
thereby uncovering potential pitfalls with implications for the
determination of the vacuum.

\section{Fluctuations in standard inflation}

In inflation, see \cite{1}, we consider the action of the scalar inflaton
field, minimally coupled to gravity:
\begin{equation}
S=\frac{1}{2}\int (\partial _{\mu }\phi \partial ^{\mu }\phi -V(\phi ))\sqrt{%
-g}~d^{4}x-\frac{1}{16\pi G}\int R\sqrt{-g}~d^{4}x  \label{2}
\end{equation}%
One assumes the background to be a homogenous isotropic Friedmann universe
with zero spatial curvature. In comoving coordinates $y$ and comoving time $%
\tau $, the metric reads ${ds}^{2}=a^{2}(\tau )\left( {d\tau }^{2}-{\delta }%
_{ij}d{y}^{i}d{y}^{j}\right) $. The perturbations of the metric tensor can
be decomposed into scalar, vector and tensor modes according to their
transformation properties under spatial coordinate transformations on the
constant-time hypersurfaces, namely $ds^{2}=ds_{S}^{2}+ds_{V}^{2}+ds_{T}^{2}$%
, where:
\begin{eqnarray}
ds_{S}^{2} &=&a^{2}(\tau )\left( (1+2\Phi )d\tau ^{2}-2\partial
_{i}Bdy^{i}d\tau -[(1-2\Psi )\delta _{ij}+2\partial _{i}\partial
_{j}E]dy^{i}dy^{j}\right)   \label{4} \\
ds_{V}^{2} &=&a^{2}(\tau )\left( d\tau ^{2}+2V_{i}dx^{i}d\tau -[\delta
_{ij}+W_{i,j}+W_{j,i}]dx^{i}dx^{j}\right)   \label{vector} \\
ds_{T}^{2} &=&a^{2}(\tau )\left( d\tau ^{2}-[\delta
_{ij}+h_{ij}]dx^{i}dx^{j}\right)   \label{tensor}
\end{eqnarray}%
This generalizes the decomposition of vector fields into a curl and a
gradient field. Here, $\Phi ,B,\Psi $ and $E$ are scalar fields, $V_{i}$ and
$W_{i}$ are 3-vector fields satisfying $V_{i,i}=W_{i,i}=0$ and $h_{ij}$ is a
symmetric three-tensor field satisfying $h_{i}^{i}=0=h_{ij}{}^{,j}$. The
inflaton field \ $\phi (\mathbf{y},\tau )$ fluctuates about its spatially
homogeneous background $\phi (\mathbf{y},\tau )=\phi _{0}(\tau )+\delta \phi
(\mathbf{y},\tau )$, where $\phi _{0}(\tau )$ is the homogenous part of the
scalar field that is driving the background expansion and the perturbation
is assumed small: $|\delta \phi |\ll \phi _{0}$. In standard inflation,
vector fluctuations are not amplified by the expansion but it should be
interesting to reconsider if this still holds true in inflation with a
minimum length. Here, we will focus on scalar and tensor fluctuations.

\subsection{Scalar perturbations}

It is the quantum fluctuations of the intrinsic curvature $\Re $ which are
thought to have seeded what later became the dominant perturbations in the
CMB. The intrinsic curvature $\Re $, which is gauge invariant, can be
expressed as
\begin{equation}
\Re =-\frac{a^{\prime }}{a}\frac{\delta \phi }{\phi _{0}^{\prime }}-\Psi ,
\end{equation}%
The prime denotes differentiation with respect to the conformal time $\tau $%
. Expanding the action to second order yields
\begin{equation}
S_{S}^{(1)}=\frac{1}{2}\int d\tau ~d^{3}\mathbf{y}~z^{2}\left( (\partial
_{\tau }\Re )^{2}-\delta ^{ij}~\partial _{i}\Re \partial _{j}\Re \right)
\label{sims}
\end{equation}%
for the action of $\Re ,$where%
\begin{equation}
z=\frac{a\phi _{0}^{\prime }}{\alpha },~~~~\alpha =a^{\prime }/a  \label{z}
\end{equation}%
Clearly, the very simplest formulation that one can give for the action of
the field $\Re $ is given in Eq. (\ref{sims}). The minimalist approach to
dealing with ordering ambiguities therefore requires one to start from this
formulation of the action when introducing $\hbar $ and the minimum length $%
\Delta x_{min}$ while not introducing any ambiguous terms by hand. This was
indeed tacitly the route taken in work that introduced the minimum length
into inflationary QFT \cite{7,20,10,11}.

In the vast literature on standard inflationary theory, however, a slight
reformulation of the action is usually preferred as the starting point for
quantization. Namely, one often introduces an auxiliary field variable, $u$,
through
\begin{equation}
u=-z\Re =a\left( \delta \phi +\frac{\phi _{0}^{\prime }\Psi }{\alpha }%
\right)   \label{R}
\end{equation}%
whose dynamics follows from the action:
\begin{equation}
S_{S}^{(2)}=\frac{1}{2}\int d\tau d^{3}\mathbf{y}\left( {\left( \partial
_{\tau }u\right) }^{2}-\delta ^{ij}~{\partial }_{i}u~{\partial }_{j}u+\frac{%
z^{\prime \prime }}{z}u^{2}\right)   \label{6}
\end{equation}%
As long as we do not introduce a minimum length, the two actions $S_{S}^{(1)}
$ and $S_{S}^{(2)}$ are equivalent. More precisely, they differ by a
boundary term:
\begin{equation}
S_{S}^{(1)}-S_{S}^{(2)}=\int d\tau ~d^{3}\mathbf{y}~\frac{d}{d\tau }\left(
\frac{z^{\prime }}{z}u^{2}\right)   \label{6b}
\end{equation}%
The reason why one often prefers to quantize starting from the action $%
S_{S}^{(2)}$ rather than from the action $S_{S}^{(1)}$ is that $S_{S}^{(2)}$
possesses no overall time-dependent factor, and this gives it the appearance
of an action of a free field theory on flat space. Its only nontrivial
aspect is that the field $u(\mathbf{y},\tau )$ has a time-varying ``mass'' $%
z^{\prime \prime }/z$. The similarity to a Minkowski space theory suggests
that in this formulation the field can be quantized in the same way that one
would quantize a field on flat space. This suggests that one can identify
the vacuum state in the same way as one does in the case of Minkowski space
theories. Concretely, the Euler Lagrange field equation reads:
\begin{equation}
\hat{u}^{\prime \prime }-{\nabla }^{2}\hat{u}-\frac{z^{\prime \prime }}{z}%
\hat{u}=0.  \label{13}
\end{equation}%
The momentum conjugate to $u(\mathbf{y},\tau )$ is given by $\pi (\mathbf{y}%
,\tau )=\frac{\partial \mathcal{L}_{s}}{\partial u^{\prime }}=u^{\prime }(%
\mathbf{y},\tau )$. To quantize, one promotes $u$ and $\pi $ to operators, $%
\hat{u}$ and $\hat{\pi}$, which satisfy canonical commutation relations on
hypersurfaces of constant $\tau $:
\begin{equation}
\lbrack \hat{u}(\tau ,\mathbf{y}),\hat{u}(\tau ,\mathbf{y^{\prime }})]=[\hat{%
\pi}(\tau ,\mathbf{y}),\hat{\pi}(\tau ,\mathbf{y^{\prime }})]=0  \label{11}
\end{equation}%
\begin{equation}
\lbrack \hat{u}(\tau ,\mathbf{y}),\hat{\pi}(\tau ,\mathbf{y^{\prime }}%
)]=i\delta ^{3}(\mathbf{y}-\mathbf{y^{\prime }})  \label{12}
\end{equation}%
Employing the plane wave expansion
\begin{equation}
\hat{u}(\tau ,\mathbf{y})=\int \frac{d^{3}\mathbf{k}}{(2\pi )^{3/2}}\left[
u_{k}(\tau )\hat{a}_{\mathbf{k}}e^{i\mathbf{k}\cdot \mathbf{y}}+u_{k}^{\ast
}(\tau )\hat{a}_{\mathbf{k}}^{\dag }e^{-i\mathbf{k}\cdot \mathbf{y}}\right]
\label{14}
\end{equation}%
the fields will obey the commutation relations Eqs. (\ref{11},\ref{12}) if
the operators $\hat{a}_{\mathbf{k}}$ obey the Fock commutation relations $[%
\hat{a}_{\mathbf{k}},\hat{a}_{\mathbf{k^{\prime }}}]=[\hat{a}_{\mathbf{k}%
}^{\dag },\hat{a}_{\mathbf{k^{\prime }}}^{\dag }]=0,[\hat{a}_{\mathbf{k}},%
\hat{a}_{\mathbf{k^{\prime }}}^{\dag }]=i\delta ^{3}(\mathbf{k}-\mathbf{%
k^{\prime }})$ and if the mode functions $u_{k}$ obey the Wronskian
condition:
\begin{equation}
u_{k}^{\ast }\frac{du_{k}}{d\tau }-u_{k}\frac{du_{k}^{\ast }}{d\tau }=-i.
\label{18}
\end{equation}%
Further, the field equation will be obeyed if the number-valued functions $%
u_{k}(\tau )$ obey the mode equation:
\begin{equation}
u_{k}^{\prime \prime }+\left( k^{2}-\frac{z^{\prime \prime }}{z}\right)
u_{k}=0.  \label{15}
\end{equation}%
At this point, initial conditions must be chosen for the solution of Eq. (%
\ref{15}). This choice is crucial because it implies the identification of
the vacuum state and this affects all predictions of the theory.
Intuitively, one expects that if a mode can be followed back to when its
proper wavelength was infinitesimally short then one sees the mode when it
was virtually unaffected by curvature, i.e. here by the expansion. One
should therefore be able to identify the correct solution of the mode
equation at those early times, which then sets the initial conditions of the
mode for all time. Indeed, in Eq. (\ref{15}), one observes that $z^{\prime
\prime }/z\rightarrow 0$ at early times, $\tau \rightarrow -\infty $, i.e.
when the mode's proper wave length was arbitrarily short. In this limit, Eq.
(\ref{15}) formally turns into $u_{k}^{\prime \prime }+k^{2}u_{k}=0$ which
is the zero mass wave equation for a Minkowski space theory. For such a
theory the correct solution of the wave equation is known and one proceeds,
therefore, to impose
\begin{equation}
u_{k}(\tau )\rightarrow \frac{1}{\sqrt{2k}}~e^{-ik\tau }\mbox{~~~~~~for~~}%
\tau \rightarrow -\infty   \label{19}
\end{equation}%
as the initial condition for Eq. (\ref{15}). This identifies the initial
vacuum of each mode as the incoming lowest energy vacuum. The mathematical
problem for calculating $u$ is now well-posed. Finding $u$ then yields the
mode function for the intrinsic curvature $\Re =-u/z$ and from it we finally
obtain the observationally relevant power spectrum $P_{S}^{1/2}(k)$, of the
intrinsic curvature's quantum fluctuations after horizon crossing:
\begin{equation}
P_{s}^{1/2}(k)=\sqrt{\frac{k^{3}}{2\pi ^{2}}}\left| \Re _{k}\right| {%
\Big\vert}_{\frac{k}{aH}{\ll }1}  \label{20}
\end{equation}%
To conclude: before introducing a minimum length the actions $S_{S}^{(1)}$
and $S_{S}^{(2)}$ differ merely by a boundary term. Thus, re-expressing the
mode equation Eq. (\ref{15}) that followed from $S_{S}^{(2)}$ in terms of
the intrinsic curvature yields
\begin{equation}
\Re _{\tilde{k}}^{\prime \prime }+\frac{2z^{\prime }}{z}\Re _{\tilde{k}%
}^{\prime }+k^{2}\Re _{\tilde{k}}=0  \label{sfic}
\end{equation}%
which is of course the same mode equation that one obtains as Euler Lagrange
equation directly from the action $S_{S}^{(1)}$. The rationale for taking
the detour via the action $S_{S}^{(2)}$ is that this route exhibits a
similarity with QFT on Minkowski space, which suggests a particular choice
of initial condition and thus of the vacuum.

\subsection{Tensor perturbations}

\label{secfour} The situation for the tensor modes $h$ is similar. Their
dynamics is determined by expanding the Einstein-Hilbert action to second
order:
\begin{equation}
S_{T}^{(1)}=\frac{m_{Pl}^{2}}{64\pi }\int d\tau d^{3}\mathbf{y}~a^{2}(\tau
)~\partial _{\mu }h^{i}{}_{j}~\partial ^{\mu }h_{i}{}^{j}  \label{42}
\end{equation}%
The aim is to calculate the spectrum of the quantum fluctuations of $h$
after horizon crossing. This spectrum should become empirically testable
through measurements of the CMB's $B$-polarization spectrum, the first
measurements of which may come from the upcoming PLANCK satellite telescope.

It is clear that the action $S_T^{(1)}$ is of precisely the same form as $%
S_S^{(1)}$, up to constants and the replacement of $z(\tau)$ by $a(\tau)$.
This means that it is possible to reformulate also the tensor action to give
it the appearance of a Minkowski space theory with variable mass term,
thereby obtaining a criterion for picking the initial conditions. Therefore,
instead of quantizing directly from $S_T^{(1)}$, one often prefers to
introduce the re-scaled variable, $P^{i}{}_{j}$
\begin{equation}
P^{i}{}_{j}(y)=\sqrt{\frac{m_{Pl}^{2}}{32\pi}}~a(\tau )h^{i}{}_{j}(y)
\label{43}
\end{equation}%
whose dynamics follows from the action:
\begin{equation}
S_{T}^{(2)}=\frac{1}{2}\int d\tau d^{3}\mathbf{y}\left(\partial _{\tau
}P_{i}{}^{j}\partial ^{\tau }P^{i}{}_{j}-\delta ^{rs}{\partial }%
_{r}P_{i}{}^{j}{\partial }_{s}P^{i}{}_{j}+\frac{a^{\prime \prime }}{a}%
P_{i}{}^{j}P^{i}{}_{j}\right)  \label{44}
\end{equation}%
Analogously to the case of scalar fluctuations, the two actions $S_T^{(1)},
S_T^{(2)}$ merely differ by a term which is a total time derivative
\begin{equation}
\triangle S_{T}=S_{T}^{(2)}-S_{T}^{(1)}= \frac{32\pi}{m_{Pl}^2}\int d\tau
d^{3}\mathbf{y}\left( \alpha P_{i}{}^{j}~P^{i}{}_{j}\right) ^{\prime }
\label{45}
\end{equation}%
and therefore lead to the same equation of motion.

\noindent One proceeds by decomposing $P^{i}{}_{j}$ into its Fourier
components
\begin{equation}
P^{i}{}_{j}=\sum\limits_{\lambda =+,\times }\int \frac{d^{3}\mathbf{k}}{%
(2\pi )^{3/2}}~p_{\mathbf{k},\lambda }(\tau )~\epsilon ^{i}{}_{j}(\mathbf{k}%
;\lambda )~e^{i\mathbf{k}\cdot \mathbf{y}}  \label{46}
\end{equation}%
where $\epsilon ^{i}{}_{j}(\mathbf{k};\lambda )$ is the polarization tensor,
satisfying the conditions: $\epsilon _{ij}=\epsilon _{ji},~~\epsilon
^{i}{}_{i}=0,~~k^{i}\epsilon _{ij}=0$ and $\epsilon ^{i}{}_{j}(\mathbf{k}%
;\lambda )\epsilon ^{j\ast }{}_{i}(\mathbf{k};\lambda ^{\prime })=\delta
_{\lambda \lambda ^{\prime }}$. There are two independent polarization
states, usually denoted $\lambda =+,\times $. It is convenient to choose $%
\epsilon _{ij}(-\mathbf{k};\lambda )=\epsilon _{ij}^{\ast }(\mathbf{k}%
;\lambda )$ which implies that $p_{\mathbf{k},\lambda }=p_{-\mathbf{k}%
,\lambda }^{\ast }$. The action for tensor perturbations then takes the
form:
\begin{equation}
S_{T}^{(2)}=\sum\limits_{\lambda =+,\times }\int d\tau d^{3}\mathbf{k}%
~\left( \left( \partial _{\tau }\left| p_{\mathbf{k},\lambda }\right|
\right) ^{2}-\left( k^{2}-\frac{a^{\prime \prime }}{a}\right) \left| p_{%
\mathbf{k},\lambda }\right| ^{2}\right)   \label{48}
\end{equation}%
To quantize, one promotes $p_{\mathbf{k},\lambda }$ to an operator $\hat{p}_{%
\mathbf{k},\lambda }$ and expands it in terms of creation and annihilation
operators, $\hat{p}_{\mathbf{k},\lambda }=p_{k}(\tau )\hat{a}_{\mathbf{k}%
,\lambda }+p_{k}^{\ast }(\tau )\hat{a}_{\mathbf{k},\lambda }^{\dagger }$ to
obtain  the wave equation
\begin{equation}
p_{k}^{\prime \prime }+\left( k^{2}-\frac{a^{\prime \prime }}{a}\right)
p_{k}=0,  \label{50}
\end{equation}%
for the mode functions $p_{k}(\tau )$ (omitting the index $\lambda $).
Analogously to scalar fluctuations, also the $p_{k}(\tau )$ must obey the
Wronskian condition (\ref{18}). As for scalar modes, the similarity with the
zero mass Minkowski space wave equation at early times suggests to impose
the initial condition that the field takes the form given in (\ref{19}) for $%
k/aH\rightarrow \infty $. The mathematical problem is then well defined and $%
p$ can be calculated. From $p$ one obtains the tensor mode $h^{i}{}_{j}=%
\sqrt{\frac{32\pi }{m_{Pl}^{2}a^{2}}}p^{i}{}_{j}$ and finally the spectrum
of tensor quantum fluctuations $h_{k}$ after horizon crossing:
\begin{equation}
P_{T}^{1/2}=\sqrt{\frac{k^{3}}{2\pi ^{2}}}\left| h_{k}\right| {\Big\vert}_{%
\frac{k}{aH}\ll 1}  \label{51}
\end{equation}%
To summarize, as in the case of scalar fluctuations, one starts quantization
from the action $S_{T}^{(2)}$ so as to exploit the similarity with Minkowski
space QFT for identifying the initial conditions and thus the vacuum state
for tensor modes.

\section{Inflation with the minimum length uncertainty relation}

The minimum length uncertainty principle was first introduced into inflation
in \cite{7}. One starts by implementing the uncertainty relations in first
quantization through modifications of the canonical $x,p$ commutation
relations, as in Eq. (\ref{1}). The first quantization commutation relations
then carry over to quantum field theory. Note that since momentum space is
unaffected by the minimum length uncertainty relations the field commutators
in momentum space remain unchanged by the procedure.

This program was carried out explicitly \cite{7} for an action of the form
of the tensor action $S_{T}^{(1)}$. This showed how $\beta >0$ generalizes
the action $S_{T}^{(1)}$ to a new action $S_{T,\beta }^{(1)}$ and how the
tensor fluctuations' equation of motion changes correspondingly. Those
results immediately also translate to the case of the scalar action $%
S_{S}^{(1)}$ since this action differs merely by overall constants and by
suitably replacing $a(\tau )$ with $z(\tau )$. These equations of motion
have been further investigated in \cite{10,11}. We will explicitly list
those equations of motion below.

Our aim now is to carry out the same program for introducing the minimum
length, starting, however, from the often-used actions $S_S^{(2)}$ and $%
S_T^{(2)}$ to derive the generalized actions $S_{S,\beta}^{(2)}$ and $%
S_{T,\beta}^{(2)}$ and the correspondingly generalized equations of motion
for scalar and tensor fluctuations. We will find that the generalized
actions $S_{S,\beta}^{(1)}$ and $S_{S,\beta}^{(2)}$ as well as the
generalized actions $S_{T,\beta}^{(1)}$ and $S_{T,\beta}^{(2)}$ no longer
differ merely by boundary terms, are therefore not equivalent and lead to
slightly different equations of motion.

\subsection{Scalar fluctuations with minimum length}

The minimum length is to be introduced as a minimum proper length in the CMB
rest frame. To this end, we transform the action $S_{S}^{(2)}$ as given in
Eq. (\ref{6} from comoving coordinates $y^{i}$ and time $\tau $ to proper
coordinates $x^{i}$ and time $\tau $, where $x^{i}=a(\tau )y^{i}$. Since the
transformation is time-dependent, the chain rule leads to a nontrivial
transformation of the derivative $\partial _{\tau }$ on fields and we
obtain:
\begin{equation}
S_{S}^{(2)}=\int d\tau \frac{d^{3}\mathbf{x}}{2a^{3}}\Bigg\{\left[ \left(
\partial _{\tau }+\frac{a^{\prime }}{a}\sum\limits_{i=1}^{3}\partial
_{x^{i}}x^{i}-\frac{3a^{\prime }}{a}\right) u\right] ^{2}-a^{2}\sum%
\limits_{i=1}^{3}(\partial _{x^{i}}u)^{2}+\frac{z^{\prime \prime }}{z}u^{2}%
\Bigg\}  \label{23}
\end{equation}%
We identify $-i\partial _{x^{i}}$ as the momentum operator, $\mathbf{P}^{i}$%
, and $x^{i}$ as the position operator $\mathbf{X}^{i}$. These operators are
defined on a Hilbert space of fields (not states) with:
\begin{eqnarray}
\left( u_{1},u_{2}\right)  &=&\int d^{3}\mathbf{x}~u_{1}^{\ast }(x)u_{2}(x)
\notag  \label{24} \\
\mathbf{X}^{i}{u}(x) &=&x^{i}{u}(x)  \notag \\
\mathbf{P}^{i}u(x) &=&-i\partial _{x^{i}}u(x).
\end{eqnarray}%
The fields thus form a Hilbert space representation of the commutation
relations:
\begin{equation}
\lbrack \mathbf{X}^{i},\mathbf{P}^{j}]=i\delta ^{ij},~~[\mathbf{X}^{i},%
\mathbf{X}^{j}]=0,~~[\mathbf{P}^{i},\mathbf{P}^{j}]=0  \label{26}
\end{equation}%
This merely expresses the fact that the canonical commutation relations of
first quantization are present also in second quantization. For example, in
quantum field theory the $\hbar $ of the Fourier factor $e^{ixp/\hbar }$
directly derives from the $\hbar $ in the commutation relations of first
quantization. Of course, the operators $\mathbf{X}^{i}$ and $\mathbf{P}^{j}$
no longer possess a simple interpretation as observables. We see from Eqs. (%
\ref{6},\ref{23}) that under the time-dependent mapping from comoving to
proper positions, the chain rule makes the action of $\partial _{\tau }$ on
fields in comoving coordinates transform into a new action on fields in
proper coordinates, namely $\partial _{\tau }\rightarrow A(\tau )$, where:
\begin{equation}
A(\tau )=\left( \partial _{\tau }+i\frac{a^{\prime }}{a}\sum\limits_{i=1}^{3}%
\mathbf{P}^{i}\mathbf{X}^{i}-3\frac{a^{\prime }}{a}\right)   \label{ab1}
\end{equation}%
Using the operators $\mathbf{X}$ and $\mathbf{P}$ we can write the action (%
\ref{23}) in representation-independent form
\begin{equation}
S_{S}^{(2)}=\int \frac{d\tau }{2a^{3}}\left( \left( u,A^{\dag }(\tau )A(\tau
)u\right) -a^{2}\left( u,\mathbf{P}^{2}u\right) +\frac{z^{\prime \prime }}{z}%
\left( u,u\right) \right) ,  \label{25}
\end{equation}%
meaning that  (\ref{25}) is true without referring to the position
representation, momentum representation or any other representation of the
fields. Following \cite{7}, our strategy now is to maintain this
representation-independent form of the action while introducing a cutoff by
modifying the underlying position-momentum commutation relations (\ref{26}).
The modified commutation relations should break neither translation nor
rotation invariance and should introduce a finite minimum position
uncertainty $\Delta x_{min}$ in all three position variables. It has been
shown \cite{ak-nosc} that all such commutation relations must have the
following form
\begin{equation}
\lbrack \mathbf{X}^{i},\mathbf{P}^{j}]=i\left( \frac{2\beta p^{2}}{\sqrt{%
1+4\beta p^{2}}-1}~\delta ^{ij}+2\beta ~\mathbf{P}^{i}\mathbf{P}^{j}\right)
,~~[\mathbf{X}^{i},\mathbf{X}^{j}]=0,~~[\mathbf{P}^{i},\mathbf{P}^{j}]=0
\label{27}
\end{equation}%
to first order in the parameter $\beta $, which is chosen positive, see \cite%
{ak-nosc}. The minimum position uncertainty $\Delta x_{min}$ in every
coordinate is given by
\begin{equation}
\Delta x_{min}=\frac{\sqrt{\beta }}{2}\left( 1+d/2\right) ^{1/4}\left( \sqrt{%
1+d/2}+1\right)
\end{equation}%
where $d$ is the number of space dimensions, see \cite{kema}. Here $d=3$, so
that $\Delta x_{min}\approx 1.62~\sqrt{\beta }$. Correspondingly, $\sigma
\approx 1.62~\sqrt{\beta }H$, where $H$ is the Hubble parameter. A
convenient Hilbert space representation of the modified commutation
relations (\ref{27}), is given by
\begin{eqnarray}
\mathbf{X}^{i}\psi (\rho ) &=&i\partial _{\rho ^{i}}\psi (\rho )  \label{rp}
\\
\mathbf{P}^{i}\psi (\rho ) &=&\frac{\rho ^{i}}{1-\beta \rho ^{2}}\psi (\rho )
\label{27a}
\end{eqnarray}%
with the scalar product:
\begin{equation}
\left( \psi _{1},\psi _{2}\right) =\int_{\rho ^{2}<\beta ^{-1}}d^{3}\rho
~\psi _{1}^{\ast }(\rho )\psi _{2}(\rho )  \label{29}
\end{equation}%
Thus, the operator $A(\tau )$ changes as the minimum length is introduced,
i.e. when $\beta >0$. The action for scalars, Eq. (\ref{25}), then also
changes to become:
\begin{equation}
S_{S,\beta }^{(2)}=\int d\tau \int_{\rho ^{2}<\beta ^{-1}}d^{3}\rho ~\frac{1%
}{2a^{3}}\Bigg\{\left| \left( \partial _{\tau }-\frac{a^{\prime }}{a}\frac{%
\rho ^{i}}{1-\beta \rho ^{2}}\partial _{\rho ^{i}}-\frac{3a^{\prime }}{a}%
\right) u\right| ^{2}-\frac{a^{2}\rho ^{2}|u|^{2}}{(1-\beta \rho ^{2})^{2}}+%
\frac{z^{\prime \prime }}{z}|u|^{2}\Bigg\}  \label{30}
\end{equation}%
The presence of $\rho $ derivatives means that the $\rho $ modes are
coupled. Conveniently, in the new variables $(\tilde{\tau},\tilde{k})$
\begin{eqnarray}
\tilde{\tau} &=&\tau ,  \notag  \label{31} \\
\tilde{k}^{i} &=&a\rho ^{i}e^{-\beta \rho ^{2}/2}
\end{eqnarray}%
the $\tilde{k}$ modes decouple. To see this, note that:
\begin{equation}
\partial _{\tau }-\frac{a^{\prime }}{a}\frac{\rho ^{i}}{1-\beta \rho ^{2}}%
~\partial _{\rho ^{i}}=\partial _{\tilde{\tau}}.  \label{32}
\end{equation}%
We will use the common index notation $\bar{u}_{\tilde{k}}$ for the
decoupling modes. The $\tilde{k}$ modes only coincide with the usual
comoving modes on large scales, i.e., only for small $\rho ^{2}$. This
means that the precisely comoving $k$ modes, obtained by scaling
$k^{i}=ap^{i}$, decouple only at large distances;  at distances close to
the cutoff scale they couple. The action now takes the decoupled form
(i.e. there are no $\tilde{k}$ derivatives):
\begin{equation}
S_{S,\beta }^{(2)}=\int d\tilde{\tau}\int_{\tilde{k}^{2}<a^{2}/e\beta }d^{3}%
\tilde{k}~a^{-6}~\frac{\kappa }{2}\left( \left| \left( \partial _{\tau }-%
\frac{3a^{\prime }}{a}\right) \bar{u}_{\tilde{k}}(\tau )\right| ^{2}-\mu
\left| \bar{u}_{\tilde{k}}\right| ^{2}+\frac{z^{\prime \prime }}{z}\left|
\bar{u}_{\tilde{k}}\right| ^{2}\right)   \label{34}
\end{equation}%
with the functions $\mu $ and $\kappa $ are defined through
\begin{eqnarray}
\mu (\tau ,\tilde{k}) &=&-\frac{a^{2}}{\beta }\frac{W({-\beta \tilde{k}%
^{2}/a^{2}})}{(1+W({-\beta \tilde{k}^{2}/a^{2}}))^{2}}  \label{36} \\
\kappa (\tau ,\tilde{k}) &=&\frac{e^{-\frac{3}{2}W({-\beta \tilde{k}%
^{2}/a^{2}})}}{1+W({-\beta \tilde{k}^{2}/a^{2}})}
\end{eqnarray}%
where $W$ is the Lambert $W$ function (see e.g. \cite{19}), which is defined
so that $W(x)e^{W(x)}=x$. As expected, each comoving mode $\tilde{k}$ has a
starting time, $\tau _{c}$, namely the time at which $a(\tau _{c})=\sqrt{%
e\beta \tilde{k}^{2}}$, which is when $\rho ^{2}=1/\beta $, which is when
the mode's proper wave length is the cutoff length. The equation of motion
that follows from the action $S_{S,\beta }^{(2)}$ is:
\begin{equation}
\bar{u}_{\tilde{k}}^{\prime \prime }+\left( \frac{\kappa ^{\prime }}{\kappa }%
-6\frac{a^{\prime }}{a}\right) \bar{u}_{\tilde{k}}^{\prime }+\left( \mu -3~%
\frac{\kappa ^{\prime }a^{\prime }}{\kappa a}-3\left( \frac{a^{\prime }}{a}%
\right) ^{\prime }+9\left( \frac{a^{\prime }}{a}\right) ^{2}-\frac{z^{\prime
\prime }}{z}\right) \bar{u}_{\tilde{k}}=0  \label{37}
\end{equation}%
The equation of motion contains a number of terms that involve the scale
factor $a$ and appears rather complicated. This is not a consequence of the
introduction of the minimum length. Instead, it is merely due to our choice
of variables. To see this, note first that the functions $\mu $ and $\kappa $
are simpler in the variables $\tau $ and $\rho _{i}$:
\begin{eqnarray}
\mu (\tau ,\rho ) &=&\frac{a^{2}\rho ^{2}}{(1-\beta \rho ^{2})^{2}}
\label{35} \\
\kappa (\rho ) &=&\frac{e^{3\beta \rho ^{2}/2}}{1-\beta \rho ^{2}}
\label{karo}
\end{eqnarray}%
Thus, as the cutoff is removed, $\beta \rightarrow 0$, we have that $\mu
\rightarrow k^{2}$ and $\kappa \rightarrow 1$. The action  (\ref{34}) thus
turns into a conventional-looking action, except for an overall factor of $%
a^{-6}$. The many terms of $a$ and $a^{\prime }$ in the equation of motion  (%
\ref{37}) trace back to this pre-factor $a^{-6}$ in the action (\ref{34}).
The occurrence of the factor $a^{6}$ might be surprising since we had
started with the action $S_{S}^{(2)}$ as given in (\ref{6}), which of course
does not possess a time-dependent pre-factor. The reason for the occurrence
of this pre-factor is that the operations of Fourier transforming and of
scaling do not commute: We did not directly Fourier transform the original
action (\ref{6}) from comoving positions to comoving momenta, as is usually
done. Instead, we first scaled the comoving position coordinates to proper
coordinates (where we introduced the minimum length), then Fourier
transformed to proper momenta, and finally scaled to comoving momenta. The
field variable $\bar{u}_{\tilde{k}}$ therefore differs from the usual field
variable $u_{\tilde{k}}$ by a factor of $a^{-3}$:
\begin{equation}
{u}_{\tilde{k}}=a^{-3}\bar{u}_{\tilde{k}}  \label{38}
\end{equation}%
In this commonly used field variable, the action $S_{S,\beta }^{(2)}$ for
scalar fluctuations, Eq. (\ref{34}), then takes the more familiar-looking
form
\begin{equation}
S_{S,\beta }^{(2)}=\int d\tilde{\tau}\int_{\tilde{k}^{2}<a^{2}/e\beta }d^{3}%
\tilde{k}~\frac{1}{2}\kappa \left( {u}_{\tilde{k}}^{\prime \ast }{u}_{\tilde{%
k}}^{\prime }-\left( \mu -\frac{z^{\prime \prime }}{z}\right) {u}_{\tilde{k}%
}^{\ast }{u}_{\tilde{k}}\right)   \label{ffa}
\end{equation}%
and also yields the equation of motion in a simpler form:
\begin{equation}
~~~~~~~~~~~~~~~~~~~~~~~~~~~~~~~~~~~~~~~~~~~~~{u}_{\tilde{k}}^{\prime \prime
}+\frac{{\kappa }^{\prime }}{{\kappa }}{u}_{\tilde{k}}^{\prime }+\left( \mu -%
\frac{z^{\prime \prime }}{z}\right) {u}_{\tilde{k}}=0,~~~~~~~~~~~~\left( %
\mbox{derived from }S_{S,\beta }^{(2)}\right)   \label{39}
\end{equation}%
Note that the introduction of the minimum length did leave us with a
time-dependent pre-factor $\kappa (\tau ,\tilde{k})$ in the action Eq. (\ref%
{ffa}), a fact that we will revisit. The mode equation (\ref{39})
generalizes Eq. (\ref{15}) in the presence of the minimal length cutoff
when starting from the action $S_{S}^{(2)}$. We need to add that the
canonical commutation relations between $u_{\tilde{k}}$ and its conjugate
momentum, $\pi _{\tilde{k}}={\kappa }u_{k}^{\prime }$, namely
\begin{equation}
\lbrack u_{\tilde{k}},\pi _{\tilde{k}^{\prime }}]=i\delta ^{3}(\tilde{k}-%
\tilde{k}^{\prime }),  \label{Wronskian1}
\end{equation}%
require that the solutions to equation (\ref{Wronskian1}) also obey the
slightly generalized Wronskian condition
\begin{equation}
u_{\tilde{k}}(\tau )u_{\tilde{k}}^{\ast ^{\prime }}(\tau )-u_{\tilde{k}%
}^{\ast }(\tau )u_{\tilde{k}}^{\prime }(\tau )=i{\kappa }^{-1}.
\label{Wronskian2}
\end{equation}%
Expressing the equation of motion (\ref{39}) in terms of the intrinsic
curvature, $\Re =-u/z$, we obtain:
\begin{equation}
~~~~~~~~~~~~~~~~~~~~~~~~~~~~~~~~~~{\Re }_{\tilde{k}}^{\prime \prime }+\left(
\frac{{\kappa }^{\prime }}{{\kappa }}+\frac{2z^{\prime }}{z}\right) {\Re }_{%
\tilde{k}}^{\prime }+\left( \mu +\frac{z^{\prime }{\kappa }^{\prime }}{z{%
\kappa }}\right) {\Re }_{\tilde{k}}=0~~~~~~~~~\left( \mbox{derived from }%
S_{S,\beta }^{(2)}\right)   \label{exa}
\end{equation}%
It is straightforward to show that the wave equation and Wronskian equation
reduce to the usual wave equation (\ref{15}) and Wronskian condition (\ref%
{18}) in the limit $\beta \rightarrow 0$, i.e. when the minimum length
cutoff is removed.

To summarize, we calculated the generalization of the action $S_{S}^{(2)}$
to the action $S_{S,\beta }^{(2)}$ and found the corresponding equation of
motion (\ref{39}).

Let us now compare with the result of introducing the minimum length
uncertainty relation into the action $S_{S}^{(1)}$ to obtain $S_{S,\beta
}^{(1)}$. We read off from \cite{7} that the action $S_{S,\beta }^{(1)}$
yields the wave equation:
\begin{equation}
~~~~~~~~~~~~~~~~~~~~~~~~~~~~~~~~~~~~~~~~~~~~~~~~~~~~~\Re _{\tilde{k}%
}^{\prime \prime }+\left( \frac{{\kappa }^{\prime }}{{\kappa }}+\frac{%
2z^{\prime }}{z}\right) \Re _{\tilde{k}}^{\prime }+\mu \Re _{\tilde{k}%
}=0~~~~~~~~~~~~~~~~~~~~\left( \mbox{derived from }S_{S,\beta }^{(1)}\right)
\label{newscalar2}
\end{equation}%
It is expressed in terms of the intrinsic curvature. In order to better
compare with the equation of motion (\ref{39}) which followed from $%
S_{S,\beta }^{(2)}$, we rewrite Eq. (\ref{newscalar2}) in terms of the field
variable $u$, to obtain:
\begin{equation}
~~~~~~~~~~~~~~~~~~~~~~~~~~~~~~~~~~~~~~~~~~~~~~~{u}_{\tilde{k}}^{\prime
\prime }+\frac{{\kappa }^{\prime }}{{\kappa }}{u}_{\tilde{k}}^{\prime
}+\left( \mu -\frac{z^{\prime \prime }}{z}-\frac{z^{\prime }}{z}\frac{{%
\kappa }^{\prime }}{{\kappa }}\right) {u}_{\tilde{k}}=0~~~~~~~~~~~~\left( %
\mbox{derived from }S_{S,\beta }^{(1)}\right)   \label{newscalar1}
\end{equation}%
The results of \cite{7} also show that this field $u_{\tilde{k}}$ satisfies
the same Wronskian condition (\ref{Wronskian2}). Clearly, the equations of
motion (\ref{39}) and (\ref{newscalar1}) differ and we will need to
investigate the origin and extent of the difference.

\subsection{Tensor fluctuations with minimum length}

\label{modten} From the case of scalar fields we find the corresponding two
actions $S_{T,\beta }^{(1)}$ and $S_{T,\beta }^{(2)}$ for tensor
perturbations, namely by inserting suitable constants and by replacing
occurrences of $z$ by $a$. For $\beta =0$ the two actions are of course
equivalent, differing merely by a boundary term. For $\beta >0$, however, we
find that they yield slightly different equations of motion:
\begin{eqnarray}
~~~~~~~~~~~~~~~~~~~~~~~~~~~~~{h}_{\tilde{k}}^{\prime \prime }+\left( \frac{{%
\kappa }^{\prime }}{{\kappa }}+\frac{2a^{\prime }}{a}\right) {h}_{\tilde{k}%
}^{\prime }+\mu {h}_{\tilde{k}}=0 &&~~~~~~~~~~~~~\left( \mbox{derived from }%
S_{T,\beta }^{(1)}\right)   \label{56} \\
{p}_{\tilde{k}}^{\prime \prime }+\frac{{\kappa }^{\prime }}{{\kappa }}{p}_{%
\tilde{k}}^{\prime }+\left( \mu -\frac{a^{\prime \prime }}{a}-\frac{%
a^{\prime }}{a}\frac{{\kappa }^{\prime }}{{\kappa }}\right) {p}_{\tilde{k}%
}=0 &&~~~~~~~~~~~~~\left( \mbox{derived from }S_{T,\beta }^{(1)}\right)
\label{ten1me} \\
{h}_{\tilde{k}}^{\prime \prime }+\left( \frac{{\kappa }^{\prime }}{{\kappa }}%
+\frac{2a^{\prime }}{a}\right) {h}_{\tilde{k}}^{\prime }+\left( \mu +\frac{%
a^{\prime }{\kappa }^{\prime }}{a{\kappa }}\right) {h}_{\tilde{k}}=0
&&~~~~~~~~~~~~~\left( \mbox{derived from }S_{T,\beta }^{(2)}\right)
\label{54} \\
~~~~~~~~~~{p}_{\tilde{k}}^{\prime \prime }+\frac{{\kappa }^{\prime }}{{%
\kappa }}{p}_{\tilde{k}}^{\prime }+\left( \mu -\frac{a^{\prime \prime }}{a}%
\right) {p}_{\tilde{k}}=0 &&~~~~~~~~~~~~~\left( \mbox{derived from }%
S_{T,\beta }^{(2)}\right)   \label{53}
\end{eqnarray}

\subsection{Origin of the differences in the mode equations}

In order to trace the inequivalence of the obtained equations of motion, we
begin by noting that we encountered an ordering ambiguity in Eqs. (\ref{ab1},%
\ref{27}) when modifying the commutation relations: consider the formal
position and momentum operators in the operator $A(\tau )$. We could have
used the first quantization's canonical commutation relations to
arbitrarily re-order the positions and momenta. Clearly, it does matter,
however, whether we do this before or after we change the first
quantization's commutation relations. In this way, for example by adding
terms of the form $(xp-px-i\hbar )f(x,p)$ before changing the commutation
relations, we could have introduced into the action arbitrary terms that
vanish as the minimum length is set to zero, i.e. as $\beta \rightarrow
0$.

Of course, in any theory that generalizes quantum field theory by
introducing a minimum length parameter one can guess Hamiltonians and
actions etc. only up to terms that vanish as the minimum length parameter
vanishes - much like quantum Hamiltonians can be guessed from classical
Hamiltonians only up to terms that vanish as $\hbar \rightarrow 0$. We
encountered essentially an instance of Dirac's observation that quantization
removes degeneracy. As in the case of quantization, the minimalist approach
to dealing with the ambiguity is to bring the action into a simple form and
to not use the ambiguity to introduce by hand any such terms that would
vanish as the minimum length is set to zero. This was the approach tacitly
adopted in \cite{7} and we here also adopted the same minimalist approach
when we introduced the minimum length into $%
S_{S}^{(1)},S_{S}^{(2)},S_{T}^{(1)}$ and $S_{T}^{(2)}$. We then found that
actions $S_{S,\beta }^{(1)}$ and $S_{S,\beta }^{(2)}$ (and similarly $%
S_{T,\beta }^{(1)},S_{T,\beta }^{(2)}$) yield differing equations of motion.
How could this happen, given that the two actions $S_{S}^{(1)}$ and $%
S_{S}^{(2)}$ (and similarly $S_{T}^{(1)},S_{T}^{(2)}$) are equivalent?

We already indicated that the answer traces back to the fact that the
actions in the two formulations of types $S^{(1)}$ and $S^{(2)}$ are
equivalent only up to a boundary term. After the minimum length is
introduced these terms are no longer boundary terms. To see that this is the
case, consider the scalar actions $S_{S,\beta }^{(2)}$ in (\ref{30}) as
expressed in terms of the field $u_{\tilde{k}}$. Note that it possesses in
its integration measure a time-dependent factor
\begin{equation}
\int d\tau ~d^{3}\tilde{k}~{\kappa }(\tau ,\tilde{k})
\end{equation}%
If we remove the minimum length, $\beta \rightarrow 0$, we obtain of course $%
{\kappa }\rightarrow 1$. In the case $\beta >0$, however, if
\begin{equation}
\int d\tau d^{3}\tilde{k}~\Delta \mathcal{L}=\int d\tau d^{3}\tilde{k}~\frac{%
d}{d\tau }f(\tau )
\end{equation}%
is a negligible boundary term arising from a total time derivative, then in
the presence of the minimum length uncertainty relation
\begin{equation}
\int d\tau d^{3}\tilde{k}~{\kappa }(\tau ,\tilde{k})~\Delta \mathcal{L}=\int
d\tau d^{3}\tilde{k}~{\kappa }(\tau ,\tilde{k})~\frac{d}{d\tau }f(\tau )
\end{equation}%
is not a boundary term. The same phenomenon occurs for tensor fluctuations:
the two actions which are normally equivalent because differing merely by
the total time derivative $\triangle S_{T}$ given in Eq. (\ref{45} now yield
different equations of motion. Indeed, as expected, when the minimal length
is introduced the two actions differ by:
\begin{equation}
S_{T,\beta }^{(2)}-S_{T,\beta }^{(1)}=\int d\tau d^{3}\mathbf{\tilde{k}}%
~\left( a^{\prime }ah_{\mathbf{\tilde{k}}}^{2}\right) ^{\prime }~{\kappa }%
(\tau ,\tilde{k})  \label{58}
\end{equation}%
The integrant is generally not a total time derivative due to the presence
of the function ${\kappa }(\tau ,\tilde{k})$.

\subsection{Comparison of the equations of motion}

The equations of motion that arise from the actions of type $S^{(1)}$ and
type $S^{(2)}$ differ merely in their ``mass terms'', i.e. in the terms that
multiply the undifferentiated field. The mass terms differ by the terms
\begin{equation}
c(\tau ,\tilde{k})=\frac{z^{\prime }\kappa ^{\prime }}{z\kappa },~~~~%
\mbox{and}~~~~d(\tau ,\tilde{k})=\frac{a^{\prime }\kappa ^{\prime }}{a\kappa
}
\end{equation}%
in the scalar and tensor cases respectively. Let us now compare, for
example, the equations of motion for tensors - the situation is completely
analogous for scalars. How significant is the term $d(\tau ,\tilde{k})=\frac{%
a^{\prime }\kappa ^{\prime }}{a\kappa }$ by which the two equations  (\ref%
{ten1me}) and (\ref{53}) differ? The term $d$ competes with the two other
``mass'' terms, $a^{\prime \prime }/a$ and $\mu $. In Fig.1, the magnitudes
of the terms $d,\frac{a^{\prime \prime }}{a}$ and $\mu $ are compared by
plotting the logarithm of their absolute values against conformal time. We
chose the de Sitter background with a realistic Planck length to Hubble
length ratio of $\sigma =10^{-5}$. In de Sitter space all $\tilde{k}$ modes
evolve in the same way and we arbitrarily chose $\tilde{k}=1$.
\begin{figure}[t]
\includegraphics[angle=0, scale=0.6]{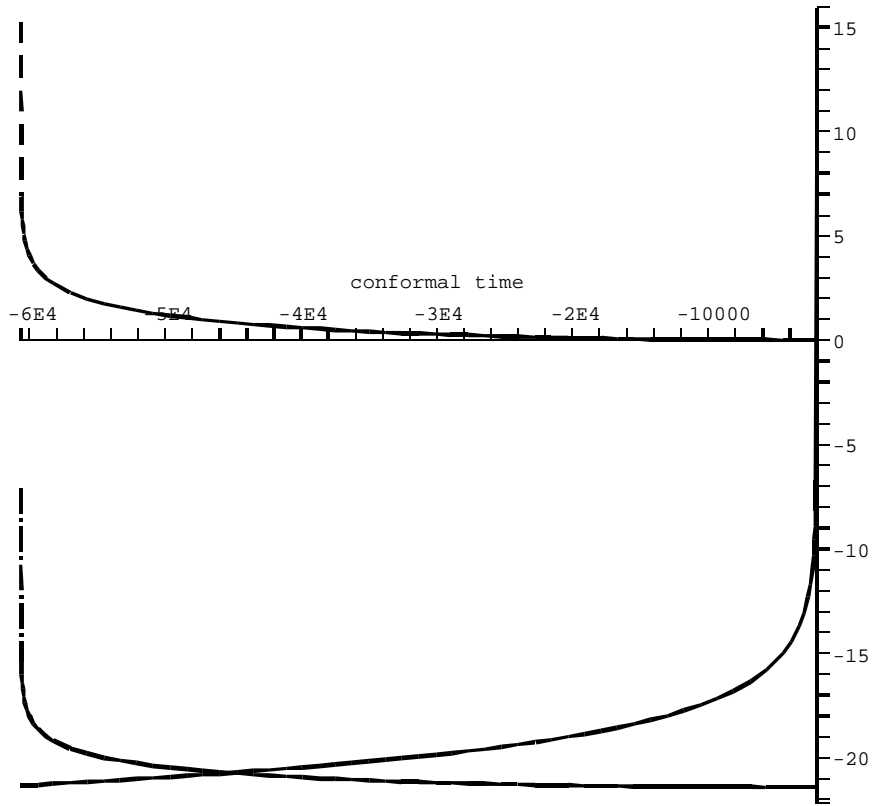}
\caption{Comparison of the three ``mass terms'' $\ln (\protect\mu )$
(dashed), $\ln (|d|)$ (dotted) and $\ln (a^{\prime \prime }/a)$ (solid)
versus conformal time. The term $d$ of the ambiguity is dominated by $%
\protect\mu $ and $a^{\prime \prime }/a$ throughout the evolution.}
\end{figure}
The curves start at the creation time $\tau _{c}$ of the mode and end at
future infinity $\tau =0$. There are three distinct phases in a mode's
evolution:

A) In the initial phase close to the creation time, the behavior of the
differential equation is dominated by the terms $\mu $ and $d$ which both
appear to diverge. (The function $a^{\prime }/a$ is regular at $\tau _{c}$
since the creation time of a particular mode is not a special time for the
scale factor.) The behavior at creation time is crucial, however, because
this is where the initial conditions for the mode are to be set by some
suitable criterion for choosing the vacuum state of the system. We need to
determine, therefore, the relative magnitudes of $d$ and $\mu $ as $\tau
\rightarrow \tau _{c}^{+}$. To this end, let us consider $\mu $ as a
function of the inverse proper wavelength $\rho $, as in Eq. (\ref{karo}).
This shows that $\mu $ is indeed divergent when the proper wavelength $%
1/\rho $ approaches the minimum wavelength $\sqrt{\beta }$, i.e. at the
creation time $\tau _{c}$ (Eq. (\ref{karo}) also shows that $\mu \rightarrow
k^{2}$ at late times, as expected). Now a straightforward calculation yields
for the ratio of the functions $d$ and $\mu $:
\begin{equation}
\frac{d(\tau ,\tilde{k})}{\mu (\tau ,\tilde{k})}~=~-{\sigma }^{2}\left( 5+3\,%
\mathit{W}(-{k}^{2}{\sigma }^{2}{\tau }^{2})\right)   \label{oo}
\end{equation}%
The range of the Lambert W function is the finite interval $[-1,0]$. Thus,
since $\mu $ is divergent at $\tau _{c}$, also $d$ is divergent at $\tau _{c}
$. However, Eq. (\ref{oo}) also implies that at all times the term $d$ is
much smaller than the term $\mu $, namely by a factor $\sigma ^{2}$, up to a
pre-factor of order one. Since $\mu $ dominates $d$ by a factor of order $%
\sigma ^{2}$, criteria for determining the mode's initial condition at $\tau
_{c}$ are generally only correspondingly weakly affected by the presence or
absence of the term $d$.

B) The initial period is followed by an adiabatic period in which all three
terms $\mu,d$ and $a^{\prime\prime}/a$ are slowly varying. This phase lasts
until horizon crossing. In order to estimate the effect of the term $d$ on
the time evolution in this phase we compare the oscillation frequencies $%
\Omega_{d}= \sqrt{\mu-a^{\prime\prime}/a-d}$ and $\Omega_{0}=\sqrt{%
\mu-a^{\prime\prime}/a}$ in the adiabatic phase with and without the term $d$
respectively. During the adiabatic phase we can neglect the term $%
a^{\prime\prime}/a$ and we can set $\mu\approx \tilde{k}^2$ to obtain: $%
\Omega_d/\Omega_0= (1-d/(2\mu))+{\mathcal{O}}(d^2/\mu^2)$, i.e. $\Omega_d
\approx \Omega_0 (1 + b \sigma^2)$ and thus $\Delta \Omega \approx b \tilde{k%
} \sigma^2$ where $b$ is of order one. In principle, this frequency shift
alters the number $N$ of oscillations during the duration of the second
phase. In order to estimate $N$, we note that the $\tilde{k}$ mode is
created at the time $\tau_c=-\frac{1}{H\tilde{k}\sqrt{e\beta}}$; the
duration, $T$, of the second phase is comparable to this and so is of the
order of $T\approx\frac{1}{H\tilde{k}\sqrt{e\beta}}$~. Recall that $\sigma
\approx 1.62 ~\sqrt{\beta} H$, which implies $T\approx 1.62/(\tilde{k} \sigma%
\sqrt{e})$. Thus, in the course of the adiabatic second phase, the presence
or absence of the term $d$ implies approximately $N$ more or fewer
oscillations, where
\begin{equation}
N \approx \Delta \Omega~T \approx \frac{1.62~b~\tilde{k} ~\sigma^2}{\tilde{k}%
\sigma\sqrt{e}} \approx \sigma
\end{equation}
which is far less than a single oscillation.

C) The last period, from horizon crossing to the infinite future $%
\tau\rightarrow 0$ is not clearly resolved in our plot. It is the period
when the term $a^{\prime\prime}/a$ diverges and entirely dominates the $\mu$
and $d$ terms since they stay finite.

\section{Conclusions}

While the framework of quantum field theory is well-tested down to distances
of about $10^{-18}m$, it is generally expected that there are corrections
due to quantum gravity when approaching the Planck length of about $10^{-35}m
$ which may well constitute a fundamental smallest length in nature.

If, therefore, there exists a finite minimum wavelength then, during
inflation, comoving modes are continually being created. Initially, a new
comoving mode will evolve under the influence of Planck scale effects but it
is clear that at late times a comoving mode's equation of motion will reduce
to the usual low-energy mode equation, namely when the mode's proper
wavelength becomes much larger than the minimum length. Thus, as was pointed
out in \cite{sta}, effects of the Planck scale can propagate into the
observable low energy realm essentially only by selecting a solution of the
mode equation which at late times differs from the usually assumed solution
for the usual mode equation.

This suggests a simple technique for exploring possible effects that
Planck scale physics could have on inflationary predictions for the CMB.
Assume that standard quantum field theory holds \textit{unchanged }down to
the minimum wavelength where modes are being created. Then, consider a
variety of possible initial conditions for the newly created modes by
applying candidate criteria for identifying the vacuum state. It is clear
that in the time translation invariant de Sitter case all effects reduce
to merely an overall re-normalization of the flat spectrum (if each mode's
initial condition is chosen by applying the same criterion). When the
Hubble parameter varies, however, then modes oscillate a variable number
of times before crossing the horizon. Thus, generically, a mode's
amplitude will be alternatingly large and small when crossing the horizon.
This can lead to
potentially observable characteristic oscillations in the spectrum \cite%
{greene3}. In this approach, quantum field theory is implicitly assumed to
hold unchanged down to the Planck scale and therefore Planck scale physics
is modelled so as to affect the predictions of inflation, for any given
evolution of the scale factor $a(\tau )$, merely through the initial
conditions. In any realistic model, of course, the quantum field theoretic
mode equations will be modified when approaching the Planck scale. This too
will have an effect on the number of oscillations that a mode undergoes
before horizon crossing and it will therefore contribute to the predicted
oscillations of the CMB spectra.

Here, we considered a concrete model for how quantum field theory is
modified when approaching the Planck length, namely by introducing the
minimum length uncertainty principle. The equations of motion then indeed
became modified at scales close to the Planck scale. As was shown in \cite%
{20,10,11}, the inflationary predictions for the CMB are to some extent
affected, possibly leading to observable oscillations in the fluctuation
spectra. However, while the equations of motion are known, the details of
the predictions still significantly depend on precisely which initial
condition is chosen, i.e. on the identification of the vacuum.

As yet, it is not fully understood in any model how Planck scale physics
determines the initial conditions of modes as they are being created, i.e.
when their proper wave length is the minimum length. Within our model of
spacetime, in which there is a minimum length uncertainty relation, the
problem of determining the initial conditions for new comoving modes is
further complicated by the fact that the mode equation possesses an
irregular singular point at the initial time, see \cite{7,20,10,11}. So
far, in the literature, a mathematical argument based on analyticity
\cite{20}
and a physical argument based on similarity to the Bunch Davies vacuum \cite%
{10,11} have been discussed and the implications for the CMB have been
investigated. Nevertheless, the crucial problem of determining the initial
state of modes when they emerge from the Planck scale in an expansion is
still essentially unsolved.

Therefore, we here reconsidered the conventional approach to fixing the
vacuum: introduce new variables in terms of which the action resembles
that of a Minkowski space theory with variable mass - a theory for which
the correct vacuum is known. Interestingly, we found that introducing the
minimum length into this reformulated action does not yield the same
theory - the action and the equations of motion differ slightly. This
raised the question as to the extent to which any predictions are affected
by this ambiguity.

Let us, therefore, recall the mechanism that leads to the prediction of
characteristic oscillations in the CMB spectra due to the minimum length
uncertainty relation. The modulus of the mode solution oscillates before
horizon crossing if it deviates from the adiabatic solution. In the case
of non-de Sitter inflation, modes then cross the Hubble horizon
alternatingly with small and large amplitudes and this translates into the
prediction of oscillations in the CMB spectra. The oscillations are the
more pronounced the more the solutions deviate from the usual adiabatic
vacuum in the phase preceding horizon crossing. For the minimum length
uncertainty relation an effect of order $\sigma $ appears possible
\cite{10,11}. This depends, of course, on the choice of initial condition.

On the basis of the current experimental approaches, it is clear that, at
best, only effects on the CMB spectrum that are of order $\sigma $ might
become observable over the foreground. Effects of order $\sigma ^{2}$ must
be expected to compete with numerous other effects such as those due to
backreaction and the nonlinearity of gravity. We found that in the mode
equation the term $\mu $ dominates the ambiguous term $d$ by a factor $%
\sigma ^{2}$.

On the one hand, this means that the ambiguous term's effect on the
amplitude of the minimum-length-induced characteristic oscillations in the
CMB spectra is very small. This is because any generic criterion for
choosing the modes' initial condition should be affected only to the order
of $\sigma ^{2}$. We note, however, that one might of course conceive of
special initial conditions that are arbitrarily different for the two cases.
For any generic criterion for initial conditions, the modes' deviation from
the adiabatic solution is, therefore correspondingly little affected by
whether one starts from the actions of type $S^{(1)}$ or those of the type $%
S^{(2)}$. On the other hand, this leaves open the possibility that the
phases of the predicted characteristic oscillations in the CMB spectra are
affected by the ambiguous term $d$. Indeed, the term $d$ implies a small
frequency shift for the mode in the period before horizon crossing. We
found, however, that the shift in the oscillation frequency accumulates only
to a small fraction ($\sigma \approx 10^{-5}$) of a single oscillation
(whose amplitude is itself at most of order $\sigma $) over the entire
adiabatic phase.

To summarize, choosing either the actions of type $S^{(1)}$ or the actions
of type $S^{(2)}$ generically leads to practically indistinguishable
predictions for the effect of the minimum length uncertainty relation on the
CMB. Nevertheless, the two actions $S^{(1)}$ or $S^{(2)}$ are actually
different and the much-sought-after precise physical criterion for
identifying the newly created modes's initial condition may be sensitive to
the correct choice of equation of motion. We draw the lesson, therefore,
that great care must be taken when introducing a further fundamental
constant such as a minimum length into quantum field theory: even if, as
usual, ordering ambiguities are resolved by adopting the minimalist
approach, previously harmless reformulations of the theory up to a boundary
term can lead to an actual change in the theory.

In the particular case at hand one may try to circumvent the problem by
starting from the most plausible actions, i.e. those of type $S^{(1)}$,
introducing the minimum length and only then reformulating the theory such
as to resemble a Minkowski space theory with variable mass. Even this
approach faces difficulties due to the introduction of a minimum length
scale, however. To see this, note that, usually, modes can be followed
arbitrarily far back in the past to when their proper wave length was
arbitrarily short and therefore essentially unaffected by curvature,
yielding an arbitrarily close match with Minkowski space. With a minimum
length, however, the initial conditions must be chosen when the proper
wavelength of the mode is the cutoff length. At that time, unavoidably, the
match with the Minkowski case is imprecise. In addition, a further
redefinition of the field variable followed by a suitable redefinition of
the time variable could preserve the appearance of a free field theory on
flat space with a time-dependent mass. This would modify, however, the
generator of time evolution, i.e. the Hamiltonian. This can affect
Hamiltonian-based criteria for picking the vacuum state; a similar point was
made in \cite{12}. An example of a criterion for choosing the vacuum that
does not rely on the Hamiltonian, but on field uncertainties instead, was
given in \cite{danielsson, greene3}. It should be interesting to apply this
criterion in the case of the minimum length uncertainty relation.

\section*{Acknowledgments}

The authors are thankful to W. H. Kinney and R. H. Brandenberger and R.
Easther for helpful discussions. This work was supported by the Natural
Sciences \& Engineering Research Council of Canada, CFI, OIT, PREA and the
Canada Research Chairs Program.

\end{document}